\documentclass[11pt,a4paper]{article}

\usepackage{mathpple}
\usepackage{graphicx}
\usepackage{url}
\usepackage{paralist}

\usepackage{geometry}
\usepackage{fancyhdr}
\begin{document}

\thispagestyle{plain}

\noindent \textbf{Preprint of:}\\
D. K. Gramotnev, T. A. Nieminen and T. A. Hopper\\
``Extremely asymmetrical scattering in gratings with varying
mean structural parameters''\\
\textit{Journal of Modern Optics} \textbf{49}(9), 1567--1585 (2002)

\hrulefill

\begin{center}

\textbf{\LARGE
Extremely asymmetrical scattering in gratings with varying
mean structural parameters
}

{\Large
D. K. Gramotnev, T. A. Nieminen and T. A. Hopper}

Centre for Medical, Health and Environmental Physics, School of
Physical Sciences, Queensland University of Technology, GPO Box
2434, Brisbane, QLD 4001, Australia

\begin{minipage}{0.8\columnwidth}
\section*{Abstract}
Extremely asymmetrical scattering (EAS) is an unusual type
of Bragg scattering in slanted periodic gratings with the
scattered wave (the $+1$ diffracted order) propagating
parallel to the grating boundaries. Here, a unique and
strong sensitivity of EAS to small stepwise variations
of mean structural parameters at the grating boundaries
is predicted theoretically (by means of approximate and
rigorous analyses) for bulk TE electromagnetic waves and
slab optical modes of arbitrary polarization in holographic
(for bulk waves) and corrugation (for slab modes) gratings.
The predicted effects are explained using one of the main
physical reasons for EAS---the diffractional divergence
of the scattered wave (similar to divergence of a laser
beam). The approximate method of analysis is based on
this understanding of the role of the divergence of the
scattered wave, while the rigorous analysis uses the
enhanced T-matrix algorithm. The effect of small and
large stepwise variations of the mean permittivity at
the grating boundaries is analysed. Two distinctly
different and unusual patterns of EAS are predicted
in the cases of wide and narrow (compared to a
critical width) gratings. Comparison between the
approximate and rigorous theories is carried out.
\end{minipage}

\end{center}

\section{Introduction}

Previously, it has been demonstrated that scattering in slanted,
strip-like, wide (compared to the wavelength) periodic gratings
with the scattered wave propagating parallel or almost parallel
to the front grating boundary is characterized by a strong
resonant increase in the scattered wave amplitude [1--12].
This type of scattering was called extremely asymmetrical
scattering (EAS). It has been shown to be radically different
from the conventional Bragg scattering in transmitting or
reflecting gratings [1--12]. For example, in addition to
the strong resonant increase of the scattered wave amplitude
at the Bragg frequency, scattering in the geometry of EAS may
also be characterized by additional unique strong resonances,
such as an additional exceptionally strong resonance in the
sidelobe structure at a frequency that is noticeably higher
than the Bragg frequency [11], additional unique resonance with
respect to angle of scattering if the scattered wave propagates
almost parallel to the grating boundaries [12], combinations of
strong simultaneous resonances in non-uniform gratings [7--9], etc.

It has also been shown that the diffractional divergence of the
scattered wave inside and outside the grating (similar to the
divergence of a laser beam of finite aperture) is responsible for
all these unique wave effects in the geometry of EAS [2--9,11,12].
The necessity of taking the divergence into account can be seen
from the fact that if the divergence is neglected and the scattered
wave propagates parallel to a strip-like grating (the geometry of EAS),
then this wave must be located within this grating. Thus we would have
a scattered beam with aperture equal to the grating width even if the
incident wave is a plane wave. This scattered beam must spread outside
the grating owing to diffractional divergence. For a more detailed
analysis of the role of diffractional divergence in EAS see [2--9,11,12].
Note that when the scattered wave propagates at a significant (usually
several degrees) angle with respect to the grating boundaries, the
diffractional divergence of the scattered wave can be neglected [12],
and we have conventional scattering in transmitting or reflecting gratings.

On the basis of understanding the role of the diffractional divergence
for scattering in the EAS geometry, a new powerful method of simple
analytical (approximate) analysis of this type of scattering has been
introduced and justified [2--9,11,12]. The main advantage of this
method is that it is directly applicable to the analysis of all types
of waves in different periodic gratings, including surface and guided
waves in periodic groove arrays [2--9,11,12]. Even in a complex
five-layer structure with non-collinear grating-assisted coupling,
the new method allowed accurate analytical analysis of extremely
asymmetrical coupling of two optical modes guided by neighbouring
optical slabs with a corrugated interface [2]. The analysis of the
applicability conditions for the new method, based on physical
speculations [12] and rigorous analysis of EAS [10], has demonstrated
that this method normally gives excellent agreement with the rigorous
theory for gratings with small amplitude. It is these gratings with
small amplitude that are of most interest from the viewpoint of EAS
since they result in strong resonant increase of the scattered wave
amplitude [2--9]. In addition, the new method provides excellent
insight into the physical reasons for EAS, which will allow
thoughtful selection of optimal structural parameters for future
EAS-based devices and techniques.

If required (for example, for very narrow gratings or large grating
amplitude), rigorous analysis of EAS of bulk electromagnetic waves
can also be carried out by means of one of the known numerically
stable rigorous approaches, such as an enhanced T-matrix approach
[13,14], S-matrix and R-matrix approaches [15,16], an S-matrix
approach using an oblique Cartesian system of coordinates [17],
a C method [18] (for EAS of guided modes in corrugation gratings),
etc. For example, in a recent paper [10], the enhanced T-matrix
approach was used for the rigorous analysis of EAS of bulk TE
electromagnetic waves in uniform holographic gratings with constant
mean dielectric permittivity. Note, however, that all these rigorous
numerical methods do not reveal the main physical reason for the
unique pattern of scattering in EAS---the diffractional divergence
of the scattered wave.

Since the divergence plays a crucial role for EAS, variations in
this divergence are expected to result in substantial variations
in the whole pattern of scattering. It is well known that stepwise
or gradual variations of mean structural parameters across a laser
beam (for example, in nonlinear wave propagation [19]) may result
in substantial variations in the diffractional divergence. Therefore,
it can be expected
that EAS will be unusually sensitive to even small variations of
mean structural parameters across the grating.

On the one hand, this unusual sensitivity may present a problem for
experimental observations of EAS, since variations of mean structural
parameters naturally occur during manufacturing of periodic gratings.
For example, etching or ruling processes that are used for the
fabrication of strip-like relief gratings on a surface of a planar
waveguide usually result in a small reduction of the mean thickness
of the waveguide (slab) in the region of the grating. Thus, mean
thickness of the guiding slab will experience a step-like variation
at the front and rear grating boundaries. Fabrication of holographic
gratings for bulk and guided optical waves also results in varying
mean dielectric permittivity in the grating. For example, if coherent
UV radiation is used for writing a grating in a photosensitive material,
the mean dielectric permittivity in the region of the grating will be
slightly increased compared with the regions outside the grating that
are not affected by the UV radiation. Such small variations of mean
structural parameters can normally be ignored in the case of
conventional Bragg scattering in transmission and reflection
gratings. However, in the case of EAS, these variations may
result in very significant changes in the pattern of scattering.

On the other hand, the unusual sensitivity of EAS to small variations
of mean structural parameters may be very useful, e.g. for the
development of new optical and acoustic sensors and precise
measurement techniques. For example, consider EAS of an optical
slab mode into another guided mode of the same slab in a strip-like
grating on one of the slab interfaces. Deposition of thin films or
layers of some substance onto the regions of the slab outside or
inside the grating will result in varying mean effective permittivity
of the slab at the grating boundaries. Thus, EAS can be used for the
detection of such deposited layers (thin film sensors).

However, no theoretical analysis of EAS of optical waves in non-uniform
gratings with varying mean structural parameters has been carried out
so far.

Therefore, the aim of this paper is to present theoretical (approximate)
and numerical (rigorous) analyses of EAS of optical bulk and guided
waves in gratings with step-like variations of mean dielectric
permittivity (for bulk TE waves) and mean slab thickness (for
optical guided TE and TM modes) at the front and rear grating
boundaries. The theoretical analysis will be based on the recently
developed approximate method allowing for the diffractional divergence
of the scattered wave [2--9,12]. It will be used for the investigation
of EAS of bulk and guided optical waves in holographic and surface
relief gratings, respectively. The rigorous numerical analysis will
be presented for bulk TE electromagnetic waves and will be based on
the enhanced T-matrix approach developed by Moharam et al. [13,14].
Comparison between the approximate and rigorous results will be
carried out. In particular, the unusually strong sensitivity of EAS
to small variations in the mean structural parameters at the front
and rear grating boundaries will be confirmed and analysed.

\section{Structure and methods of analysis}

\begin{figure}[htb]
\centerline{\includegraphics[width=0.5\columnwidth]{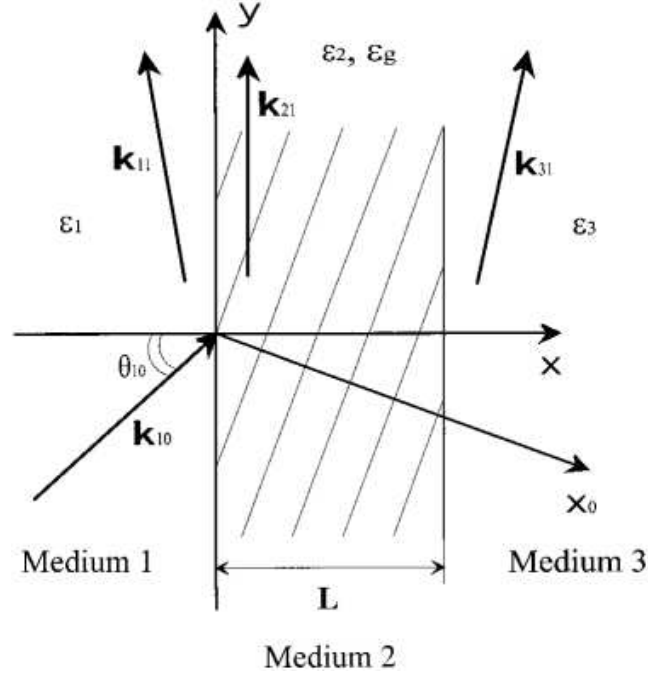}}
\caption{The geometry of EAS in a slanted holographic grating of
width $L$ with stepwise variations of the mean permittivity at the
front and rear boundaries. The vectors $\mathbf{k}_{11}$,
$\mathbf{k}_{21}$, and $\mathbf{k}_{31}$ are the wave vectors of the
scattered wave (the first diffracted order) in front, inside, and
behind the grating. For EAS of slab modes in a corrugation grating,
the plane of the figure is the plane of the slab, and $\epsilon_1$,
$\epsilon_2$, $\epsilon_3$, and $\epsilon_g$ must be replaced by the
slab thicknesses $h_1$, $h_2$, the mean slab thickness in the grating
region $h_3$, and the corrugation amplitude $\xi_g$,
respectively.}
\end{figure}

The structure analysed in this paper is presented in figure 1.
First consider bulk electromagnetic waves in a holographic grating
with sinusoidal variations of the dielectric permittivity within a
slab of thickness $L$:
\begin{equation}
\epsilon_s = \left \{ \begin{array}{l}
\epsilon_1,\\
\epsilon_2 + \epsilon_g \exp(\mathrm{i}q_xx+\mathrm{i}q_yy)
+ \epsilon_g^\ast \exp(-\mathrm{i}q_xx-\mathrm{i}q_yy) \\
\epsilon_3,
\end{array}, \textrm{for} \right \{ \begin{array}{l}
x<0,\\ 0<x<L,\\ x>L,
\end{array}
\end{equation}
where the coordinate system is shown in figure 1, $\epsilon_2$ is the
mean dielectric permittivity in the grating, $\epsilon_1$ and
$\epsilon_3$ are the dielectric permittivities of media 1 and 3
surrounding the grating (figure 1), $\epsilon_g$ is the complex
amplitude and $\mathbf{q} = (q_x,q_y)$ is the reciprocal lattice
vector of the grating that is parallel to the $x_0$-axis,
$q = 2\pi/\Lambda$, $\Lambda$ is the period and $L$ is the grating width.
The grating is assumed to be infinite along the $y$- and $z$-axes.
It is also assumed that dissipation is absent, i.e. $\epsilon_{1,2,3}$
are real and positive (the effect of dissipation on EAS is considered
in [20]). The variations of the mean permittivity at the front and rear
boundaries are given by$\Delta\epsilon_1 = \epsilon_1 - \epsilon_2$
and $\Delta\epsilon_3 = \epsilon_3 - \epsilon_2$ (these values can
obviously be real positive or negative). A TE electromagnetic wave
with the amplitude of the electric field $S_{10}$ and wave vector
$\mathbf{k}_{10}$ is incident onto the grating at an angle $\theta_{10}$
in the $xy$ plane---figure 1 (non-conical scattering).

The solution to the wave equation in the grating can be written in
the form of the expansion [13,14,21]:
\begin{equation}
E_2(x,y,t) = \sum_{n=-\infty}^{+\infty} E_{2n}(x,y,t),
\end{equation}
where
\begin{displaymath}
E_{2n}(x,y,t) = S_{2n}\exp(\mathrm{i}xk_{2nx} + \mathrm{i}yk_{2ny}
- \mathrm{i}\omega t)
\end{displaymath}
is the field in each of the inhomogeneous waves in the sum (2) with
the $x$-dependent amplitude $S_{2n}(x)$, and $k_{2nx}$ and $k_{2ny}$
are the components of the wave vectors
\begin{equation}
\mathbf{k}_{2n} = \mathbf{k}_{20} - n\mathbf{q},
\,\,\,\, (n = 0,\pm 1,\pm 2,...).
\end{equation}

If for some value of $n$ the magnitude of the wave vector $\mathbf{k}_{2n}$
is equal to $\omega\epsilon^{1/2}_2/c$, where $\omega$ is the frequency of
the wave, and $c$ is the light speed in vacuum, then the Bragg condition
is satisfied for this value of $n$, and the $n$th wave in expansion (2)
may have amplitude that is comparable with or even larger than the
amplitude of the incident wave (zeroth term in equation (2)).

In this paper it is assumed that the Bragg condition is satisfied
precisely for the first diffracted order, i.e. for $n=1$:
\begin{displaymath}
\mathbf{k}_{21} = \mathbf{k}_{20} - \mathbf{q},
\end{displaymath}
where $k_{21} = |\mathbf{k}_{21}| = \omega\epsilon^{1/2}_2/c$.
The wave vector of the scattered wave (the first diffracted order)
in the grating, $\mathbf{k}_{21}$, is parallel to the grating boundaries,
i.e. parallel to the $y$-axis (figure 1)---the geometry of EAS [1--10].

In the approximate method of analysis of EAS [2--9,11,12], only the
zeroth order (incident wave) and the first order (scattered wave) in
equation (2) are considered to be significantly non-zero (the two-wave
approximation). It has been shown that the allowance for the second-order
$x$-derivative of the scattered wave amplitude in the grating is essential
for the correct description of scattering in the geometry of EAS
[2--9,11,12]. In the approximate method [2--9,11,12], this second-order
derivative is introduced into the coupled wave equations through
consideration of the diffractional divergence of the scattered wave,
which is one of the main physical reasons for resonant wave effects
in the EAS geometry [2--12]. This approach is especially useful for
the analysis of EAS of surface and guided waves [2--9, 12], where
other methods fail to provide any reasonably simple solutions.

The approximate method developed in [2--9,12] is also directly
applicable to the case of non-uniform gratings with step-like
variations of the mean structural parameters at the grating
boundaries. For example the coupled wave equations in the grating in
the geometry of EAS can be written as [4,6,9,12]:
\begin{eqnarray}
\mathrm{d}^2S_{21}(x)/\mathrm{d}x^2 + K_0S_{20}(x) & = & 0,
\nonumber \\
\mathrm{d}S_{20}(x)/\mathrm{d}x - \mathrm{i}K_1S_{21}(x) & = & 0,
\end{eqnarray}
where
\begin{equation}
K_0 = -2k_{21}\Gamma_0\sin(\eta -\theta_{20}), \,\,\,\,
K_1 = \Gamma_1 \cos(\eta)/\cos)\theta_{20}),
\end{equation}
$\theta_{20}$ is the angle of refraction of the incident wave at the
front boundary, $\eta$ is the angle measured from the $x_0$-axis to the
wave vector of the incident wave $\mathbf{k}_{20}$ in the grating
counter-clockwise (figure 1), $\Gamma_0$ and $\Gamma_1$ are the
coupling coefficients that are determined in the conventional coupled
wave theories for non-slanted gratings with fringes parallel to the
grating boundaries.

The incident wave is partly reflected from the front grating boundary,
because the mean dielectric permittivity experiences a step-like
variation at $x = 0$. However, in the considered approximation, this
process of reflection (and refraction) is independent of scattering
inside the grating. Therefore, the reflected and refracted
waves at the front boundary $x = 0$ can be found separately from
scattering by means of the well-known Fresnel's equations for an
interface. The determined amplitude $S_{200}$ of the refracted wave
at the grating boundary $x = 0$, and the angle of refraction $\theta_{20}$
must then be used in equations (4) and (5) as the amplitude and the angle
of incidence of the incident wave at this boundary. The wave vector of the
incident wave in the grating should be taken as $\mathbf{k}_{20}$---the
wave vector of the refracted wave.

Similarly, the reflection of the incident wave at the rear boundary can
also be considered separately from the process of scattering. In this
case, the incident wave in the grating at $x = L$ is regarded as the
wave transmitted through the grating. Then the reflection from the rear
boundary will be the reflection of the transmitted wave from this
boundary, which (in the approximate method of analysis [2--9,12])
has no effect on the process of scattering.

It can also be seen that if $\epsilon_2 \ne \epsilon_1,\epsilon_3$,
then the scattered wave outside the grating should not necessarily
propagate parallel to the grating boundaries, even if this is the case
inside the grating (figure 1). In general, the solutions for the
scattered wave outside the grating are
\begin{eqnarray}
E_{11}(x,y,t) & = & A_0 \exp(\mathrm{i}k_{11y}y -
\mathrm{i}k{11x}x - \mathrm{i}\omega t), \nonumber \\
E_{31}(x,y,t) & = & B_0 \exp(\mathrm{i}k_{31y}y +
\mathrm{i}k_{31x}x - \mathrm{i}\omega t),
\end{eqnarray}
where $A_0$ and $B_0$ are the amplitudes of the scattered waves in
media 1 and 3, respectively, $k_{11y} = k_{31y} = k_{21}$,
\begin{equation}
k_{11x} = (k^2_{11} - k^2_{21})^{1/2}, \,\,\,\,
k_{31x} = (k^2_{31} - k^2_{21})^{1/2},
\end{equation}
where $k_{11} = \omega\epsilon^{1/2}_1/c$,
$k_{31} = \omega\epsilon^{1/2}_3/c$, and $k_{11x}$ and $k_{31x}$
are chosen positive real or positive imaginary. Note that in the
rigorous analysis, equations (6) are replaced by the Rayleigh
expansions of the electric field in the regions $x < 0$ and $x > L$
[13--16,18,21]. In this case the waves with amplitudes $A_0$ and $B_0$
(equations (6)) are the first diffracted orders in these expansions.

In the two-wave approximation, the boundary conditions at the grating
boundaries can then be written as [2--9,12]:
\begin{displaymath}
S_{20}|_{x=0}=S_{200}, S_{21}|_{x=0}=A_0,
(\mathrm{d}E_{21}/\mathrm{d}x)_{x=0}=(\mathrm{d}E_{11}/\mathrm{d}x)_{x=0},
\end{displaymath}
\begin{equation}
E_{21}|_{x=L}=E_{31}|_{x=L},
(\mathrm{d}E_{21}/\mathrm{d}x)_{x=L}=(\mathrm{d}E_{31}/\mathrm{d}x)_{x=L}.
\end{equation}

Note again that the refraction of the incident (transmitted) waves at
the front and rear boundaries of the grating has not been considered
when writing boundary conditions (8).

For the rigorous theory, the boundary conditions at the grating boundaries
are given in [13,14,21].

Boundary conditions (8) determine unknown wave amplitudes $A_0$, $B_0$,
and three constants of integration in the solutions to the coupled wave
equations (4) in the grating region [2--6]. These constants determine
the incident and scattered wave amplitudes inside and outside the grating
(for the explicit form of the solutions see [2--6]).

\section{Numerical results}

The numerical results of the approximate and rigorous analyses of EAS in
gratings described by equation (1) are presented in this section for bulk
TE electromagnetic waves. As has been indicated in [6--9,11,12], there are
two typical patterns of EAS, which correspond to narrow and wide gratings.
Narrow gratings are those whose widths are less than a critical width $L_c$,
wide gratings are those with widths larger than $L_c$ [6--9,11,12].
Physically, $L_c/2$ is the distance within which the scattered wave
(beam) can be spread across the grating (i.e. in the direction normal
to the wave propagation) by means of the diffractional divergence, before
being re-scattered by the grating [7--9]. Simple methods of determination
of the critical width were developed in [7, 9]. This critical width is
extremely important for scattering in the geometry of EAS. It appears to
determine numerous resonant effects in uniform and non-uniform gratings
[7-- 9,11,12]. Strong differences in the patterns of scattering will
also be demonstrated here for EAS in narrow and wide gratings with
step-like variations of the mean structural parameters.

\subsection{Narrow gratings}

\begin{figure}[htb]
\centerline{\includegraphics[width=0.7\columnwidth]{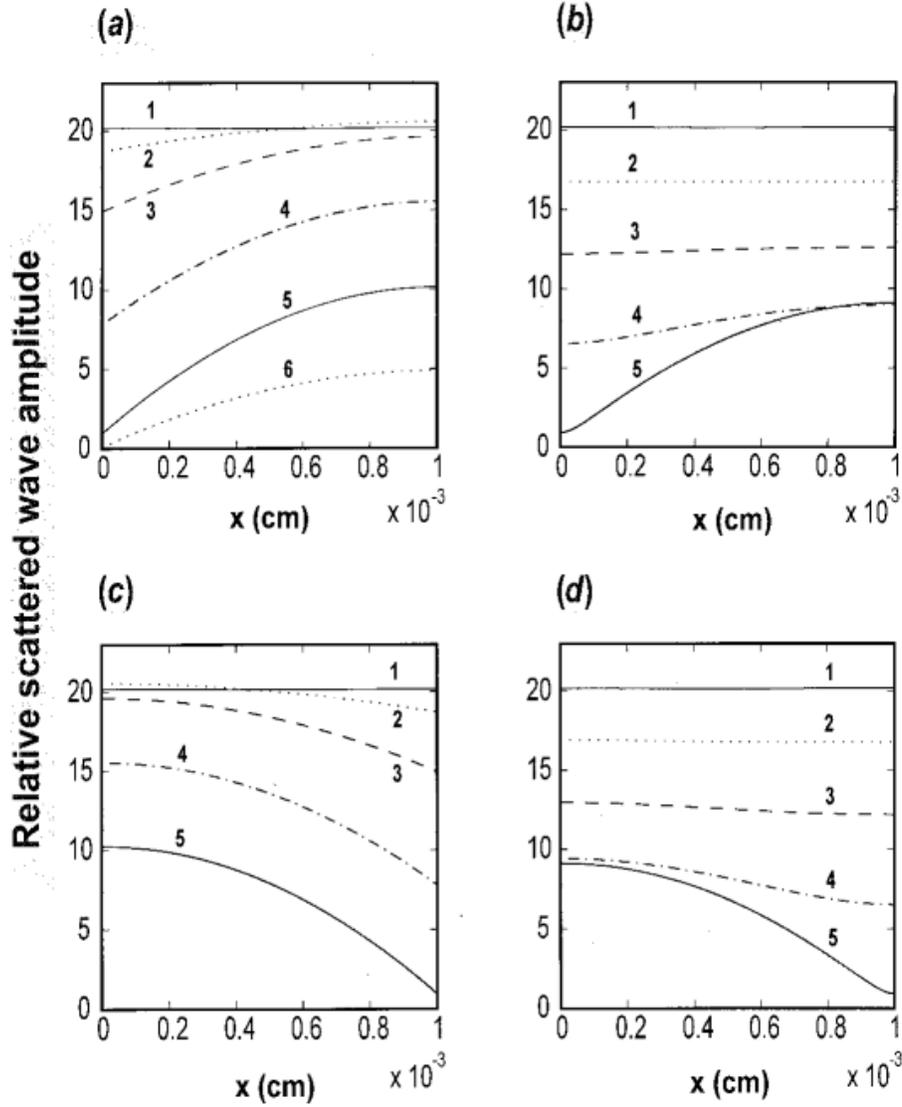}}
\caption{The $x$-dependencies of relative amplitudes $|S_{21}(x)/S_{10}|$
of the scattered bulk
TE electromagnetic wave in the holographic grating with the parameters:
$\epsilon_2=5$, $\epsilon_g=5\times 10^{-3}$, $L=10\mu$,
$\theta_{10}=\pi/4$, $\lambda(\mathrm{vaccum})=1\mu$, the grating period
($\Lambda \approx 0.584\mu$m) and the orientation of the fringes are
determined by the Bragg condition and the direction of the wave vector
$\mathbf{k}_{21}$.
(a, b) $|Delta\epsilon_1=\epsilon_1 - \epsilon_2 \ne 0$ and
$|Delta\epsilon_3=\epsilon_3 - \epsilon_2 = 0$ (variations of the mean
permittivity only at the front boundary);
(c, d) $|Delta\epsilon_1=\epsilon_1 - \epsilon_2 = 0$ and
$|Delta\epsilon_3=\epsilon_3 - \epsilon_2 \ne 0$ (variations of the mean
permittivity only at the rear boundary). Curves 1: $\Delta\epsilon_1=0$
and $\Delta\epsilon_3=0$ (no variations of the mean permittivity).
Curves 2, 3, 4, 5: (a) $\Delta\epsilon_1=-10^{-5}$, $-10^{-4}$,
$-10^{-3}$, $-10^{-1}$, respectively; (b) $\Delta\epsilon_1=10^{-5}$,
$10^{-4}$, $-0^{-3}$, $10^{-1}$; (c) $\Delta\epsilon_3=-10^{-5}$, $-10^{-4}$,
$-10^{-3}$, $-10^{-1}$; (d) $\Delta\epsilon_3=10^{-5}$,
$10^{-4}$, $-0^{-3}$, $10^{-1}$. Curve 6: $\Delta\epsilon_1=-4$,
i.e. $\epsilon_1=1$ (vacuum in front of the grating).}
\end{figure}

Figure 2 presents the dependencies of the normalized scattered wave
amplitude on the $x$-coordinate inside the holographic grating with
$L=10\mu$m, $\epsilon_2 = 5$, $\epsilon_g = 5\times 10^{-3}$,
$\lambda(\mathrm{vaccum})=1\mu$m, $\theta_{10}=45^\circ$, and step-like
variations of the mean permittivity at the front (figures 2(a), (b))
and rear (figures 2 (c), (d)) grating boundaries as indicated in the
figure caption. The Bragg condition is assumed to be satisfied precisely.
The above structural parameters correspond to the critical width
$L_c \approx 30\mu$m [7,9], and therefore, the dependencies in figure 2
are typical for narrow gratings with $L < L_c$.

The most important feature that can be seen from this figure is that the
scattered wave amplitude is unusually sensitive to small variations of
the mean permittivity at the grating boundaries. Even very small step-lik
e variations of the mean permittivity at either of the boundaries
($\Delta\epsilon_{1,3}=10^{-5}=2\times 10^{-6}\epsilon_2
\ll \epsilon_g=10^{-3}\epsilon_2$) may result in noticeable changes in
the scattered wave amplitude in the grating---compare curves 1 and 2 in
figures 2(b), (d). Note that further decrease of grating width results
in an approximately proportional increase of sensitivity of EAS to small
variations of the mean dielectric permittivity at the grating boundaries.
This is related to sharper EAS resonance in narrower gratings (if the
width is less than $L_c$) [4,5,12].

The unusually strong sensitivity of EAS to small variations of mean
structural parameters has no analogies in conventional Bragg scattering.
As has been mentioned in section 1, it can be explained by a strong
sensitivity of the diffractional divergence of the scattered wave to
small variations of structural parameters along the wave front.

Another interesting aspect is that varying mean permittivity at one of
the grating boundaries results in significant changes of the scattered
wave amplitude throughout the structure (figures 2(a)--(d)). Moreover,
if the dielectric permittivity outside the grating is larger than inside
(i.e. $\Delta\epsilon_{1,3} > 0$---see figures 2(b), (d)), then the
scattered wave amplitude is almost constant across the grating (especially
for small values of $\Delta\epsilon_{1,3}$). If
$\Delta\epsilon_{1,3} < 0$ (figures 2(a), (c)), then the dependence of the
scattered wave amplitudes on the $x$-coordinate is fairly noticeable.
However, if the grating width is reduced further (e.g. down to
$\approx 5\mu$m), then the scattered wave
amplitude becomes approximately constant across the grating for both the
cases with $\Delta\epsilon_{1,3} > 0$ and $\Delta\epsilon_{1,3} < 0$.
This is because in narrow gratings (with $L < L_c$) the diffractional
divergence is very efficient in spreading the scattered wave across the
grating (see also [6--9]). Thus, any changes in the scattered wave
(caused, for example, by varying mean structural parameters) at one
boundary of the grating are effectively felt at the other boundary
(and everywhere in the grating). Therefore, variations of the scattered
wave amplitude in the grating are smoothed out by the diffractional
divergence, and this results in only weak dependence of the scattered
wave amplitude on the $x$-coordinate (see curves 1--4 in figures
2(a)--(d)). If the grating width is decreased, the diffractional
divergence becomes even more
efficient (at shorter distances), and the $x$-dependencies of the
scattered wave amplitude become even weaker for all values of
$\Delta\epsilon_{1,3}$.

It can also be seen that if $\Delta\epsilon_{1,3} > 0$ (i.e. the
dielectric permittivity outside the grating is larger than inside),
then the sensitivity of EAS to small variations of the mean permittivity
at the grating boundaries is stronger than for $\Delta\epsilon_{1,3} < 0$
(compare curves 1--4 in figures 2(a), (c) with curves 1--4 in figures 2(b),
(d)). This is because if $\Delta\epsilon_{1,3} < 0$, then the waves
outside the grating are exponentially decaying with increasing distance
from the grating (see also section 2). These waves cannot be associated
with an energy flow towards or away from the grating. On the other hand,
if $\Delta\epsilon_{1,3} > 0$ (figures 2(b), (d)), then the scattered
waves outside the grating (according to Snell's law) are propagating
waves travelling away from the grating. The resultant energy losses from
the grating cause more significant (than in the case of
$\Delta\epsilon_{1,3} < 0$) reduction in the scattered wave amplitudes
(figures 2(a)--(d)). The smaller the grating width and/or grating amplitude
$\epsilon_g$, the stronger the difference in sensitivity of EAS to small
positive and small negative values of $\Delta\epsilon_{1,3}$.

Note that the approximate and rigorous analyses of EAS in the structures
considered give practically indistinguishable $x$-dependencies of the
scattered wave amplitude in the grating. Therefore, the dependencies
presented in figures 2(a)--(d) can equally be regarded as approximate
and rigorous (within an accuracy of $\approx 0.1$\%).

\begin{figure}[htb]
\centerline{\includegraphics[width=0.7\columnwidth]{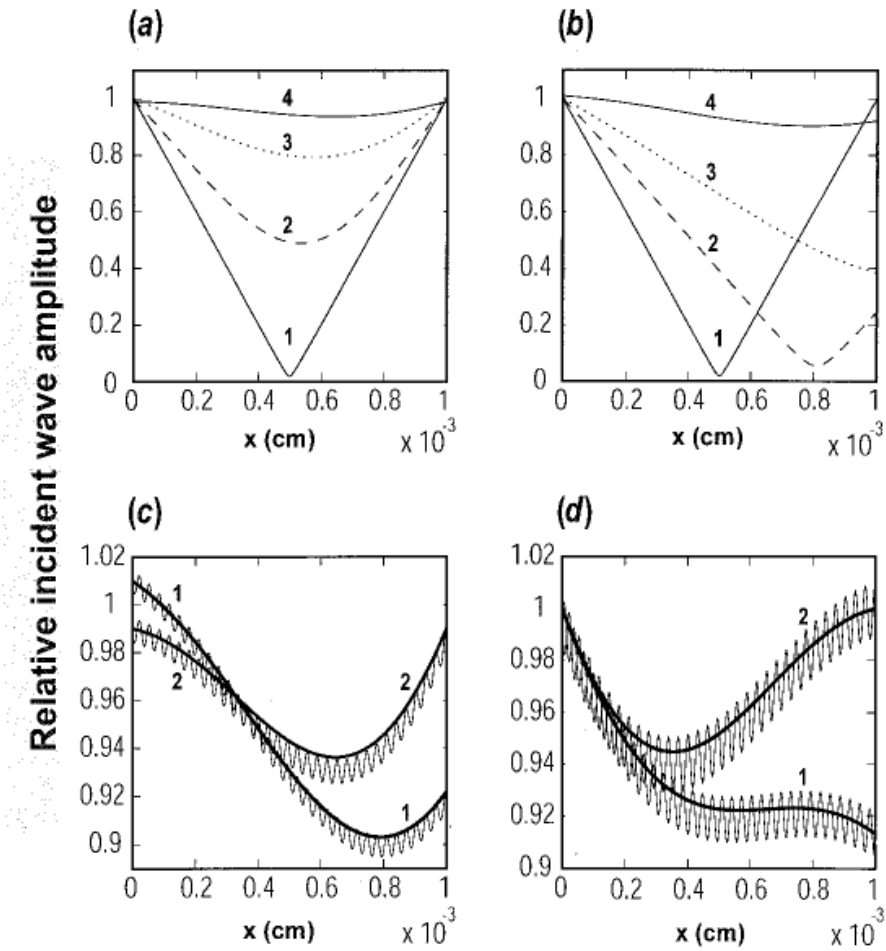}}
\caption{The $x$-dependencies of relative amplitudes $|S_{20}(x)/S_{10}|$
of the incident wave
inside the holographic grating with the same parameters as for figure 2
($\epsilon_2=5$, $\epsilon_g=5\times 10^{-3}$, $L=10\mu$m,
$\theta_{10}=\pi/4$, $\lambda(\mathrm{vacuum})=1\mu$m).
(a, b) $\Delta\epsilon_1 = \epsilon_1 - \epsilon_2 \ne 0$ and
$\Delta\epsilon_3 = \epsilon_3 - \epsilon_2 = 0$ (variations of the mean
permittivity only at the front boundary). Curves 1: $\Delta\epsilon_1=0$,
$\Delta\epsilon_3=0$. Curves 2, 3, 4: (a) $\Delta\epsilon_1 = -10^{-4}$,
$-10^{-3}$, $-10^{-1}$, respectively; (b) $\Delta\epsilon_1 = 10^{-4}$,
$10^{-3}$, $10^{-1}$. (c, d) The approximate (thick solid curves) and
rigorous (oscillating thin curves) dependencies of the relative incident
wave amplitudes in the same grating but with (c) $\Delta\epsilon_3=0$,
$\Delta\epsilon_1= 0.1$ (curves 1) and $\Delta\epsilon_1=-0.1$ (curves 2);
(d) $\Delta\epsilon_1=0$, $\Delta\epsilon_3=0.1$ (curves 1) and
$\Delta\epsilon_3=-0.1$ (curves 2).}
\end{figure}

Typical approximate dependencies of the normalized amplitude of the
incident wave on the $x$-coordinate inside a narrow grating are presented
in figures 3(a), (b) for the same structure as figures 2(a), (b), i.e. with
$\epsilon_2=5$, $\epsilon_g=5\times 10^{-3}$, $\lambda=1\mu$m,
$\theta_{10}=45^\circ$, $L=10\mu$m, and $\Delta\epsilon_3=0$ (i.e. there is
no variation of the mean permittivity at the rear grating boundary).

If the variation of the mean permittivity at the front boundary
$\Delta\epsilon_1$ < 0 (i.e. $\epsilon_1 < \epsilon_2$---figure 3(a)),
the scattered wave does not carry the energy away from the grating, and
energy conservation results in the same magnitudes of the amplitudes of
the incident wave at the front ($x=0$) and rear ($x=L$) boundaries
(figure 3(a)). If $\Delta\epsilon_1 > 0$ (i.e.
$\epsilon_1 > \epsilon_2$---see figure 3(b)), then the scattered wave at
$x < 0$ is a propagating wave carrying energy away from the grating. As
a result, the amplitude of the incident wave at the rear boundary is
smaller than at the front boundary (figure 3(b)).

If the mean permittivity varies at the rear boundary, i.e.
$\Delta\epsilon_3 \ne 0$ and $\Delta\epsilon_1 = 0$, then for small
variations of the mean permittivity ($\Delta\epsilon_{1,3}\ll\epsilon_2$)
the approximate dependencies of the incident wave amplitude in narrow
gratings are approximately the same as for the case with
$\Delta\epsilon_3 = 0$ and $\Delta\epsilon_1 \ne 0$ (figures 3(a), (b)).
The smaller the grating width, the more accurate is this statement. Note,
however, that this is not true for large variations of the mean
permittivity $\Delta\epsilon_{1,3} \approx \epsilon_2$, for which
reflection of the incident wave from the front boundary becomes noticeable.

If $\Delta\epsilon_1$ and/or $\Delta\epsilon_3$ are positive in a
narrow grating with $L < L_c$, then it is possible to choose optimal
values of $\Delta\epsilon_{1,3} \ll \epsilon_2$ such that the amplitude
of the incident wave at the rear boundary is next to zero. The smaller
the grating width, the smaller the amplitude of the incident wave at the
rear boundary can be. In this case, almost total conversion of the energy
of the incident wave into energy of the scattered wave occurs. This is
radically different from the conventional scattering in narrow gratings
with small amplitude, where scattering is very inefficient and the
scattered wave amplitude is small (as is the energy flow in it).

If $\Delta\epsilon_3=0$ and $\Delta\epsilon_1 \ne 0$, then the rigorous
$x$-dependencies of the amplitude of the zeroth diffracted order are
practically the same as the corresponding approximate dependencies of
the incident wave amplitude (figure 3(c)). The main distinctive feature
of the rigorous dependencies is the weak fast oscillations with the
period of the order of the wavelength (figure 3(c)). These oscillations
are related to boundary scattering of the scattered wave at the grating
interface $x=L$ [10]. The wave resulting from this boundary scattering
propagates in the negative $x$-direction as if it is a mirror reflected
incident wave from this boundary. Therefore, mathematically, the zeroth
term in equation (2) includes both the incident wave and the wave caused
by the boundary scattering [10]. Interference of these two waves results
in a standing wave pattern represented by fast oscillations of amplitude
of the zeroth diffracted order. The period of these oscillations,
$\lambda/(2\epsilon^{1/2}_2\cos\theta_{20})$, is in excellent agreement
with the period of oscillations of the rigorous dependencies in figure 3(c).

It is interesting that if variations of the mean permittivity occur at
the rear grating boundary, i.e. $\Delta\epsilon_3\ne 0$, then the
amplitude of oscillations of the rigorous
dependencies may be significantly larger (figure 3(d)). Increasing
$|\Delta\epsilon_3|$ results in increasing amplitude of these oscillations.
This is due to reflection of the incident wave from the rear grating
boundary with $\Delta\epsilon_3 \ne 0$. Mathematically, the reflected
wave is also included in the zeroth term in equation (2). Its amplitude
increases with increasing $|\Delta\epsilon_3|$, which results in stronger
oscillations of the resultant interference pattern.

It follows that the difference between the approximate and rigorous curves
in figure 3(d) can be reduced by including the reflected wave into the
approximate analysis. Indeed, the total electric field resulting from
the interference of the incident and reflected waves is given as
\begin{equation}
[S_{20}(x) + S_{2r}\exp(-2\mathrm{i}k_{20x}x)]
\exp(\mathrm{i}k_{20x}x + \mathrm{i}k_{10y}y - \mathrm{i}\omega t),
\end{equation}
where $S_{20}(x)$ is the $x$-dependent amplitude of the incident wave in
the grating, and $S_{2r}$ is the amplitude of the wave reflected from the
rear grating boundary (as mentioned above, $S_{2r}$ is given by the Fresnel
equations for reflection of the incident wave with the amplitude
$S_{20}(L)$ from the interface $x=L$). Wave (9) has a periodically
varying amplitude (the expression in the square brackets). If we plot
the $x$-dependencies of the magnitude of this amplitude for the gratings
corresponding to curves 1 and 2 in figure 3(d), the resultant approximate
(oscillating) curves will be very close to the rigorous oscillating curves
in figure 3(d). (Note that in the rigorous approach the term
$S_{2r} \exp(-2\mathrm{i}k_{20x}x)$ is formally included in the amplitude
$S_{20}(x)$.)

\subsection{Wide gratings}

\begin{figure}[!t]
\centerline{\includegraphics[width=0.7\columnwidth]{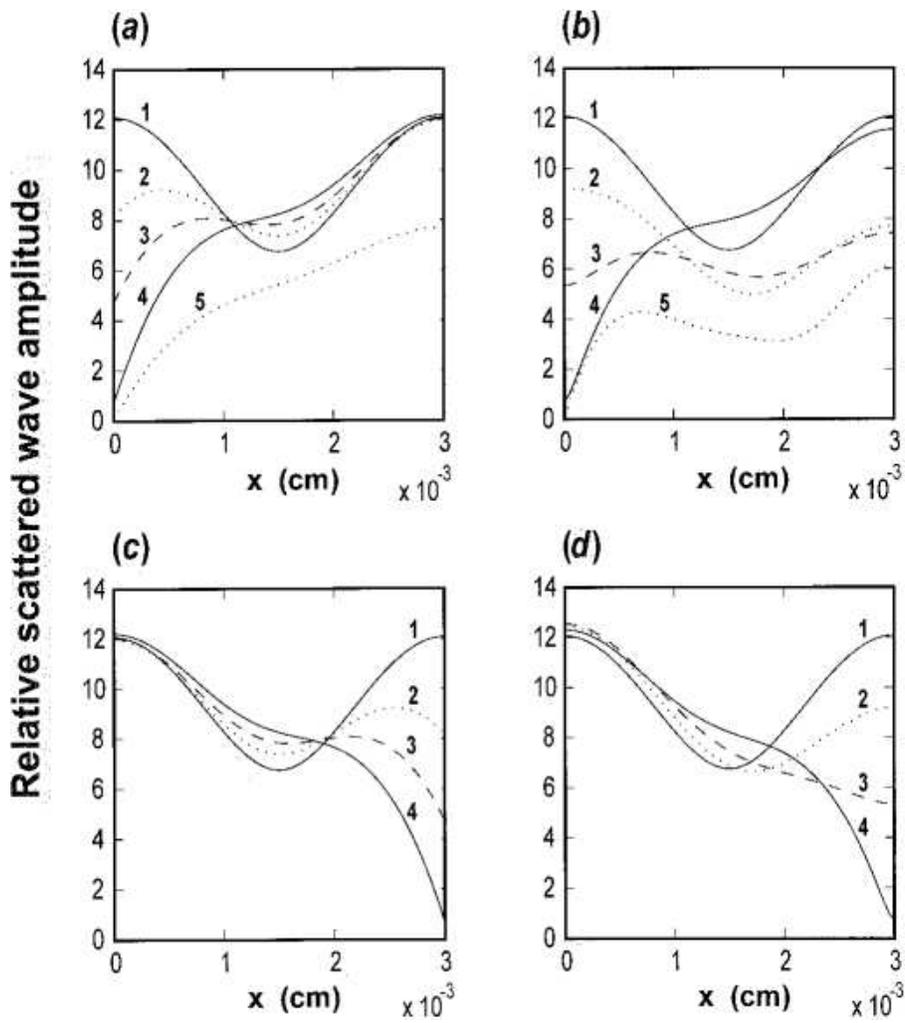}}
\caption{The $x$-dependencies of relative scattered wave amplitudes
$|S_{21}(x)/S_{10}|$ inside
the holographic grating with the same parameters as for figure 2, but with
$L\approx L_c \approx 30\mu$m. (a, b) $\Delta\epsilon_1 \ne 0$ and
$\Delta\epsilon_3 = 0$, (c, d) $\Delta\epsilon_1=0$ and
$\Delta\epsilon_3 \ne 0$. Curves 1: $\Delta\epsilon_1 = 0$ and
$\Delta\epsilon_3 = 0$. Curves 2, 3, 4: (a) $\Delta\epsilon_1 = -10^{-4}$,
$-10^{-3}$, $-10^{-1}$, respectively; (b) $\Delta\epsilon_1 = 10^{-4}$,
$10^{-3}$, $10^{-1}$; (c) $\Delta\epsilon_3 = -10^{-4}$,
$-10^{-3}$, $-10^{-1}$; (d) $\Delta\epsilon_3 = 10^{-4}$,
$10^{-3}$, $10^{-1}$. Curves 5: (a) $\Delta\epsilon_1 = -4$, i.e.
$\epsilon_1 = 1$ (vacuum in front of the grating), and (b)
$\Delta\epsilon_1 = 4.9$ (in this case the angle of incidence
$\theta_{10}=45^\circ$ is very close to the critical angle
$\approx 45.29^\circ$ for the interface $x=0$).}
\end{figure}

\begin{figure}[!t]
\centerline{\includegraphics[width=0.7\columnwidth]{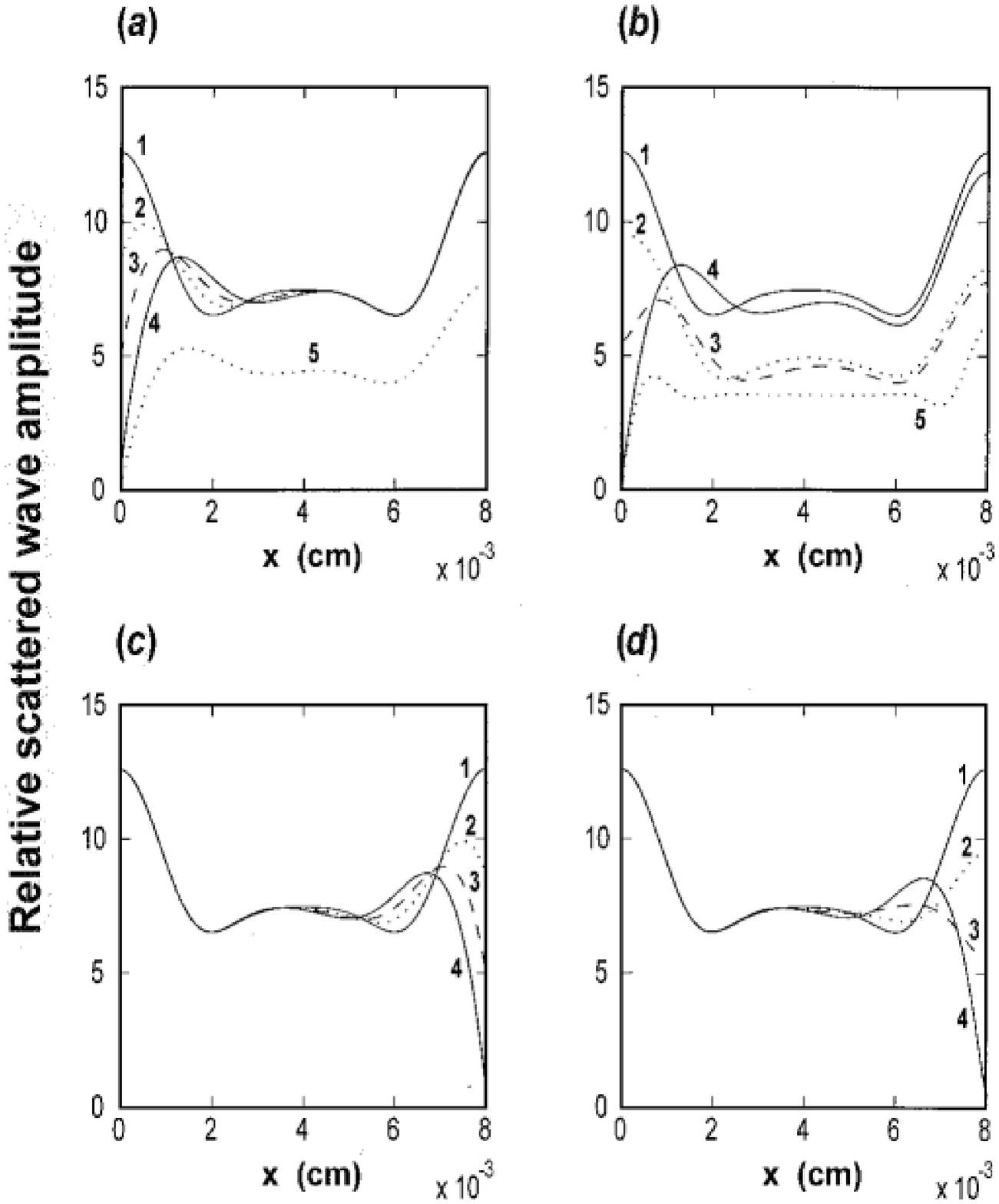}}
\caption{Same as figure 4, but for grating width
$L=80\mu\mathrm{m} > L_c \approx 30\mu$m.}
\end{figure}

If the grating width $L \ge L_c$, then the pattern of scattering changes
significantly. Typical $x$-dependencies of the scattered wave amplitude
inside the grating of critical width ($L=L_c\approx 30\mu$m) are presented
in figure 4. It can be seen that in this case the effect of varying mean
permittivity on the scattered wave amplitude tends to be localized in the
half of the grating that is adjacent to the boundary at which the variation
occurs (figure 4). This tendency becomes much more obvious if we consider
EAS in wide gratings, e.g. with $L=80\mu\mathrm{m} > L_c\approx 30\mu$m%
---figure 5.

If the variation of the mean permittivity occurs at the rear boundary
(figures 4(c), (d) and 5(c), (d)), then the effect of this variation on
the scattered wave amplitude is always localized within the distance
$\approx L_c/2$ near the rear boundary. Everywhere else in the grating,
the scattered wave amplitude is hardly affected by the varying mean
permittivity for small and large, positive and negative values of
$\Delta\epsilon_3$ (figures 4(c), (d) and 5(c), (d)). This is because the
diffractional divergence can effectively spread the scattered wave only
within distances $\approx L_c/2$ [7--9,12], and thus the effects of any
perturbations at the rear boundary (e.g. variation of the mean permittivity)
can be felt only within these distances (figures 4(c), (d) and 5(c), (d)).

If a variation of the mean permittivity occurs at the front boundary, i.e.
$\Delta\epsilon_1 \ne 0$, then for negative values of $\Delta\epsilon_1$
(figures 4(a), 5(a)) the situation is largely the same as in figures 4(c),
(d) and 5(c), (d). However, if $|\Delta\epsilon_1|$ is large (of the order
of $\epsilon_2$), then the scattered wave amplitude is affected (reduced)
everywhere in the grating---see curves 5 in figures 4(a), 5(a). This is
mainly because when $\Delta\epsilon_1$ is large and negative, the amplitude
of the incident TE wave transmitted through the boundary $x=0$ into the
grating is noticeably less than the amplitude of the
incident wave in the region $x < 0$. This must also result in a reduction
of the scattered wave amplitude everywhere in the grating.

If $\Delta\epsilon_1 > 0$ (figures 4(b), 5(b)), then increasing
$\Delta\epsilon_1$ results in a monotonous decrease of the scattered
wave amplitude near the front boundary to about zero (similar to figures
4(a), 5(a)). However, inside the grating (and at the rear boundary) the
situation is different. The scattered wave amplitude decreases with
increasing $\Delta\epsilon_1$ (curves 1--3 in figures 4(b), 5(b)), goes
through a minimum, increases back to approximately the same values as for
gratings with $\epsilon_1 = \epsilon_2 = \epsilon_3$ (see curves 1 and 4 in
figures 4(b), 5(b)), and then decreases again for large $\Delta\epsilon_1$
(curves 5 in figures 4(b), 5(b)).

The more complex behaviour of the curves in figures 4(a), (b) and 5(a),
(b) compared to the curves in figures 4(c), (d) and 5(c), (d) is mainly
related to the fact that if a perturbation occurs at the front boundary,
then the effect of this perturbation can be spread into the grating not
only by means of the diffractional divergence of the scattered wave, but
also by means of the incident wave that propagates into the grating. At
the same time, the effect of perturbations at the rear boundary can be
spread into the grating only by means of the diffractional
divergence of the scattered wave. Therefore the effect of varying mean
permittivity at the front boundary may affect the wave amplitudes everywhere
in the grating (figures 4(a), (b) and 5(a), (b)), whereas variations of the
mean permittivity at the rear boundary may affect the wave amplitudes only
within the distance $\approx L_c/2$ from the rear boundary (figures 4(c),
(d) and 5(c), (d)).

The noticeable differences between figures 4(a), 5(a) and figures 4(b),
5(b) are due to the fact that if $\Delta\epsilon_1 > 0$ (figures 4(b),
5(b)), then the scattered wave in the region $x < 0$ carries the energy
away from the grating (see also the discussions of figure 2). The
resultant energy losses from the grating can cause noticeable reduction
of the scattered wave amplitude in the whole grating (see curves 1--3
in figures 4(b), 5(b)). However, if the variation of the mean permittivity
$\Delta\epsilon_1$ is sufficiently large, then the amplitude of the
scattered wave at the front boundary is small (curves 4 in figures
4(b), 5(b)). As a result, the energy losses drastically reduce, and the
scattered wave amplitude inside the grating (not near the front boundary)
increases---compare curves 2, 3 and 4 in figures 4(b), 5(b). For a
particular value of $\epsilon_1$ (in our examples this is
$\epsilon_1 = \epsilon_{1c}=10$), the angle of incidence $\theta_{10}$
(if it is non-zero) becomes equal to the critical angle for the interface
$x=0$. In this case, the incident wave in the grating propagates parallel
to this interface, i.e. $\theta_{20} = \pi/2$. On the other hand, if the
angle of incidence in the grating $\theta_{20} \rightarrow \pi/2$,
the amplitude of the scattered wave is strongly reduced [4]. This is
the reason for
the reduction of the scattered wave amplitudes represented by curves 5
in figures 4(b) and 5(b).

If the mean permittivity varies simultaneous at both the boundaries, then
the effect of such variations is roughly the superposition of the effects
of separate variations at each of the boundaries. This statement holds
better for small variations of mean parameters.

An important general feature of EAS can be seen from the approximate
and rigorous analyses of scattering of bulk electromagnetic waves in
narrow and wide gratings. Increasing or decreasing wavelength $p$ times
with the simultaneous increasing or decreasing grating width $p$ times
leaves the amplitudes of the scattered and incident waves unchanged
(though scaled to different grating width). For example, EAS of bulk
TE waves of $\lambda = 2\mu$m in gratings with $L=160\mu$m will be
represented by exactly the same curves as in figures 5(a)--(d) with $x/2$
instead of $x$ on the horizontal axis (scaling to the two times larger
grating width). The other structural parameters are assumed to be the
same as for figure 5.

Similarly, $p$ times increase or decrease of the dielectric permittivity
in the structure (including grating amplitude and step-like variations
of the mean permittivity) with the simultaneous $p^{1/2}$ times decrease
or increase of the grating width also leaves the wave amplitudes unchanged
(though again scaled to the different grating width).

\section{EAS of guided modes}

As has been mentioned in the Introduction, the approximate method of
analysis of EAS described is directly applicable to the case of EAS of
guided optical modes of arbitrary polarization in gratings with constant
mean structural parameters [2,5,6,9,12]. Here, we will see that this is
also true for EAS of optical modes guided by a slab with varying mean
thickness at the grating boundaries. In this case, the plane of figure 1
is the plane of the slab. One of the boundaries of the slab is periodically
corrugated within the strip of width $L$ (figure 1). Thus the thickness of
the waveguide inside and outside the grating is given by the equation:
\begin{equation}
h(x_0) = \left \{ \begin{array}{l}
h_1,\\
h_2 + \xi_g f(x_0) \\
h_3,
\end{array}, \textrm{for} \right \{ \begin{array}{l}
x<0,\\ 0<x<L,\\ x>L,
\end{array}
\end{equation}
where $h_2$ is the mean thickness of the slab inside the grating region,
$h_1$ and $h_3$ are the slab thicknesses outside the grating, $\xi_g$
is the grating amplitude (corrugation amplitude), $f(x_0)$ is a periodic
function with the period $\Lambda$, $\mathrm{max}(|f(x_0)|)=1$,
and the average (over the period) value of $f(x_0)$ is equal to zero.
Dissipation is neglected and all media in contact are isotropic. The
variations of the mean slab thickness at the front and rear boundaries
are given by $\Delta h_1 = h_1 - h_2$ and $\Delta h_3=h_3-h_2$
($\Delta h_{1,3}$ can obviously be either positive or negative).

As in the case of bulk electromagnetic waves in structures described by
equation (1), the most interesting case of resonantly strong EAS of
guided modes occurs at small grating amplitudes:
\begin{equation}
|\xi_g|/\Lambda \ll 1.
\end{equation}
In this case, the approximate theory (based on the two-wave approximation
and the scalar theory of diffraction) is again expected to give highly
accurate results (see [9,10,12]).

Using speculations similar to those in section 2, one comes to the
conclusion that equations (4) also describe EAS of guided modes in
gratings with varying mean thickness at the grating boundaries. However,
the coupling coefficients $\Gamma_0$ and $\Gamma_1$ are obviously different
from those for bulk TE waves and are determined in the approximate theories
of conventional Bragg scattering in corrugation gratings with grooves
parallel to the grating boundaries [22--25]. The wave numbers for the
incident and scattered modes (e.g. $k_{11}$, $k_{21}$, $k_{20}$, etc.)
are determined by the dispersion equation for the corresponding slab modes.
The effective dielectric permittivities corresponding to these modes can be
introduced by the equations: $\epsilon_{21\mathrm{eff}}=k_{21}c/\omega)^2$,
$\epsilon_{20\mathrm{eff}}=k_{20}c/\omega)^2$, etc.

Similarly to bulk electromagnetic waves experiencing reflection and
refraction at the grating boundaries with non-zero $\Delta\epsilon_{1,3}$,
guided incident and scattered modes also interact with the step-like
variations of the mean thickness of the slab. For example, interaction
of the incident mode with a step-like variation of $h$ at the front
boundary results in generation of other guided modes in the slab and
bulk electromagnetic waves outside the slab [22--26]. In the approximate
theory, however, all these waves do not contribute to scattering in the
grating, except for one guided mode for which the Bragg condition is
satisfied. Therefore, the situation is similar to the approximate theory
of EAS of bulk electromagnetic waves in section 2. For example, from
consideration of the interaction of the incident mode with the front
boundary (by means of the mode matching theory [22--26]), one can
determine the amplitudes and the directions of propagation of the
resulting bulk and guided waves, and then regard the guided mode with
the amplitude $S_{200}$, the angle of propagation $\theta_{20}$, and the
wave vector $\mathbf{k}_{20}$ that satisfies the Bragg condition as the
incident wave at $x=0$ in the grating (inside the grating, the amplitude
of this wave is $S_{20}(x)$).

It follows that equations (4)--(7) are directly applicable to the analysis
of EAS of guided modes in corrugation gratings with step-like variations of
the mean slab thickness. In this case, however, appropriate equations have
to be used for the coupling coefficients, wave numbers, and it is assumed
that $S_{20}$, $S_{21}$, $A_0$, and $B_0$ are components of the electric
or magnetic field in the incident and scattered modes. Note also that this
approach is valid for both TE and TM guided modes.

Interaction of the scattered wave with step-like variations of the mean
thickness at the grating boundaries also results in the generation of bulk
and guided waves. This means that boundary conditions (8) (except for the
first one) have to be modified in order to include these bulk and guided
waves. This can again be done in the general case in the same way as in
the mode matching theory [22--26].

However, manufacturing gratings with small amplitude usually results in
small variations of the mean thickness of the slab. Adsorption and
deposition of thin films on the slab surface (e.g. in the region $x < 0$)
also results in only small variations of mean slab parameters (thickness
or effective dielectric permittivity). Therefore, small variations of mean
slab thickness are of most importance from the viewpoint of experimental
observation and applications of EAS.

The analysis of EAS in such structures does not require the general
approaches of the mode matching theory and can be carried out by means
of approximate boundary conditions similar to (8). This is because if
$\Delta h_{1,3} \ll h_2$, then reflection
and transformation of modes at such small stepwise variations of the
mean slab thickness can be neglected. It can be seen that in this case,
for TM modes, the boundary conditions at the grating boundaries $x=0$
and $x=L$ can be written as
\begin{displaymath}
S_{20}|_{x=0}=S_{200}, E_{21z}|_{x=0}=E_{11z}|_{x=0},
(\mathrm{d}E_{21z}/\mathrm{d}x)_{x=0}=(\mathrm{d}E_{11z}/\mathrm{d}x)_{x=0},
\end{displaymath}
\begin{equation}
E_{21z}|_{x=L}=E_{31z}|_{x=L},
(\mathrm{d}E_{21z}/\mathrm{d}x)_{x=L}=(\mathrm{d}E_{31z}/\mathrm{d}x)_{x=L}.
\end{equation}
For TE modes, the $z$-components of the electric fields should be replaced by
the $z$-components of the magnetic field. $S_{20}$ is the amplitude of one of
the components of the electric (or magnetic) field in the incident TM (or TE)
mode inside the grating region (note that if $\Delta h_1 \ll h_2$, then
$S_{200} \approx S_{10}$---the amplitude of the incident wave in the region
$x < 0$, and $\theta_{20} \approx \theta_{10}$). The incident and scattered
modes are assumed to be of arbitrary order.

It can be seen that equations (12) are exactly the same as equations (8)
with the only difference that the $z$-components of the electric or
magnetic fields in equations (12) replace the total electric field
in equations (8).

When using the boundary conditions (12), it is convenient to use effective
dielectric permittivities for the scattered modes:
$\epsilon_{11\mathrm{eff}}=k_{11}c/\omega)^2$,
$\epsilon_{21\mathrm{eff}}=k_{21}c/\omega)^2$, and
$\epsilon_{31\mathrm{eff}}=k_{31}c/\omega)^2$ in front, inside and behind
the grating, respectively. In this case, step-like variations of the mean
slab thickness at the grating boundaries are related to the corresponding
step-like variations of the mean effective permittivities. For example,
for the front boundary:
\begin{equation}
\Delta\epsilon_{11\mathrm{eff}}= \epsilon_{11\mathrm{eff}} -
\epsilon_{21\mathrm{eff}} \approx
\frac{2k_{21}c^2}{\omega^2} \left(
\frac{\mathrm{d}k_{21}}{\mathrm{d}h} \right)_{h=h_2} \Delta h_1.
\end{equation}
As mentioned above, this formula (and the similar one for the rear
boundary) is correct only if $\Delta h_{1,3} \ll h_2$. Note that we do
not have to consider step-like variations of the effective permittivity
for the incident wave. These variations can be neglected since for
$\Delta h_{1,3} \ll h_2$ we have: $S_{10}\approx S_{200}$ and
$\theta_{20}\approx\theta_{10}$ (and similar for the rear boundary),
i.e. interaction of the incident mode with the grating boundaries can
be neglected. (Note that, though being small,
$\Delta\epsilon_{11\mathrm{eff}}$ cannot be neglected owing to its
significant effect on the diffractional divergence of the scattered wave.)

For example, consider EAS of an incident TE zeroth (TE0) mode into the
scattered TE0 mode in the structure: vacuum--GaAs slab (permittivity
12.25)--AlGaAs substrate (permittivity 10.24); the mean slab thickness
within the grating region $h_2=6\times 10^{-5}$cm,
$\theta_{20}\approx\theta_{10}=\pi/4$ (for small values of
$\Delta h_1 \ll h_2$), the wavelength in vacuum
$\lambda=1.5\times 10^{-4}$cm, the corrugation is assumed to be sinusoidal,
i.e. $f(x_0)=\sin(x_0)$, and the effective permittivity
$\epsilon_{21\mathrm{eff}}\approx 11.48$. The period of the corrugation,
determined by the Bragg condition, is $\Lambda\approx 0.579\mu$m.
The wavelength of the guided TE$_0$ mode in this structure is $0.42\mu$m,
which is 92\% of the wavelength ($0.45\mu$m) in the grating used for
figures 2--5. Therefore, according to the general tendency discussed
at the end of section 3, in order to get the same dependencies for the
guided modes as in figures 2--5, we need to use the gratings of widths
$0.92L$ (where $L$ is the width in figures 2--5), and the step-like
variations of the slab thickness determined by equation (13) with
\begin{equation}
\Delta\epsilon_{1,3\mathrm{eff}}\approx
\frac{\Delta\epsilon_{1,3}}{\epsilon_2} \epsilon_{21\mathrm{eff}},
\end{equation}
where $\epsilon_2=5$ and $\Delta\epsilon_{1,3}$ are the variations of
the mean permittivity in figures 2--5. In this case, the corrugation
amplitude can be chosen ($\xi_g\approx 2\times 10^{-6}$cm) so that the
$x$-dependencies of amplitudes of the scattered and incident slab modes
in the grating are exactly the same as those given by figures 2--5 with
the replacement of $x$ by $x/0.92$ on the horizontal axis (scaling to the
new grating width).

It can be seen from equations (13) and (14) that the variations of the
mean dielectric permittivity in figures 2--5
$\Delta\epsilon_{1,3} = 10^{-5}$, $10^{-4}$, $10^{-3}$, $10^{-1}$
correspond to $\Delta h_{1,3} \approx 1.3\times 10^{-3}$nm, $0.13$\,nm,
$1.3$\,nm, 130\,nm in the considered guiding structure. Thus variations
of the mean thickness of $\approx 0.1$ monolayer may result in noticeable
variations of the scattered wave amplitude in the structure---see curve 2
in figures 2(b), (d). Obviously, on the one hand, this is a significant
complication to the experimental observation and application of EAS of
guided modes. On the other hand, this demonstrates a very high sensitivity
of EAS to conditions on slab interfaces, which may open up excellent
opportunities for the design of highly sensitive optical sensors (such
as adsorption sensors, sensors of dielectric permittivity, etc.).

If the high sensitivity of EAS to varying mean structural parameters is not
desirable, then one should use wide gratings, where the effect of varying
mean parameters at the grating boundaries is noticeable only within a half
of the critical distance---see figures 4 and 5. Also, use of larger grating
amplitudes results in reduction of the sensitivity of EAS to varying mean
parameters. Another option is to use grazing-angle scattering (GAS) [12]
rather than EAS, where the amplitude of the scattered wave may be
especially large at the middle of the grating but not at its boundaries.
There is also a possibility of compensating (at least partial) for varying
mean parameters by choosing the angle of scattering so that the scattered
wave outside the grating propagates parallel to the grating [27].

Another option for the reduction of the high sensitivity of EAS for guided
modes is to use slabs with larger thickness $h$, and permittivity that is
closer to the permittivities of the surrounding media. For example, in
the structure with a polymer slab (dielectric permittivity 2.56) on a
silica substrate (dielectric permittivity 2.22), $h=2\mu$m,
$\lambda=1.4\mu$m, $L=20\mu$m, and grating amplitude
$\xi_g=3.3\times 10^{-5}$cm we have: $\epsilon_{21\mathrm{eff}}=2.49$,
and the equivalent structure for bulk waves (having the same dependencies
of the wave amplitudes) must have $\epsilon_2=\epsilon_{21\mathrm{eff}}$,
same $\lambda$ and $L$, and $\epsilon_g=7\times 10^{-3}$. In this case,
$\Delta\epsilon_1=10^{-4}$ in the equivalent structure corresponds to a
variation in the mean thickness $\Delta h_1\approx 1.8$\,nm. These
variations of the permittivity or thickness result in $\approx 15$\%
reduction of the scattered wave amplitude from $\approx 8S_{200}$ to
$\approx 6.8S_{200}$ at the front grating boundary $x=0$. Recall that a
similar 15\% variation of the scattered wave amplitude in the GaAs
waveguide with $L=9.2\mu$m, $h=0.6\mu$m, and $\xi_g=2\times 10^{-6}$cm
is achieved at a much smaller value of
$\Delta h_1 \approx 1.3\times 10^{-2}$nm---compare curves 1 and 2
in figure 2(b).

Finally, it is important to note that in the case of EAS of surface waves
(e.g. surface acoustic waves, or surface electromagnetic waves) in
corrugation gratings, the problem with varying mean parameters hardly
exists. This is because in this case the step-like variation of the mean
level of the substrate surface inside and outside the grating does not
affect the length of the wave vector (or the wavelength) of the surface
waves. Only if, for example, outside the grating we have additional
layers on the surface, may the mean propagation parameters change and the
above-mentioned effects occur. Thus a surface wave sensor based on EAS in
periodic groove arrays is especially easy to design.

\section{Conclusions}

This paper has demonstrated yet another unique feature of scattering of
optical waves in extremely asymmetrical geometry---the unusual sensitivity
of the incident and scattered wave amplitudes to small stepwise variations
of mean structural parameters (e.g. mean dielectric permittivity or mean
waveguide thickness). This effect has no analogies in conventional Bragg
scattering. It has been explained by the high sensitivity of diffractional
divergence to small variations of mean structural parameters across an
optical beam.

Two distinct typical patterns of scattering have been described for
gratings that are narrower and wider than a grating of critical width
$L_c$ [7--9] (half of $L_c$ is equal to the distance within which the
scattered wave can be spread across the grating by means of the
diffractional divergence, before it is re-scattered by the grating [7--9]).
In particular, it has been shown that in narrow gratings (with $L < L_c$),
varying mean permittivity at either of the boundaries strongly affects the
scattered wave amplitude everywhere in the structure, whereas in wide
gratings (with $L > L_c$), the effect of varying mean parameters is
usually (but not always) significant only within the distance $L_c/2$
from the boundaries.

The analysis has been carried out by means of approximate and rigorous
methods, showing a very good agreement between them for the most
interesting cases of strong EAS in gratings with small amplitude.

The analysis of EAS of guided modes has revealed especially high
sensitivity of scattered wave amplitudes to small variations of the
mean slab thickness. This is mainly due to strong dispersion of guided
modes, i.e. strong dependence of their wave numbers on slab thickness.
This sensitivity may, on the one hand, present a complication for
experimental observation of EAS, and on the other hand, be very useful
for the application of EAS in the design of highly sensitive sensors
and measurement techniques. Several options for dealing with this
possible experimental complication are discussed.

\section*{References}

\begin{enumerate}
\item Kishino, S., Noda, A., and Kohra, K., 1972,
J. Phys. Soc. Japan., 33, 158.
\item Bakhturin, M. P., Chernozatonskii, L. A., and Gramotnev, D. K.,
1995, Appl. Opt., 34, 2692.
\item Gramotnev, D. K., 1995, Phys. Lett. A, 200, 184.
\item Gramotnev, D. K., 1997, J. Physics D, 30, 2056.
\item Gramotnev, D. K., 1997, Opt. Lett., 22, 1053.
\item Gramotnev, D. K., and Pile, D. F. P., 1999, App. Opt., 38, 2440.
\item Gramotnev, D. K., and Pile, D. F. P., 1999, Phys. Lett. A, 253, 309.
\item Gramotnev, D. K., and Nieminen, T. A., 1999,
J. Opt. A: Pure Appl. Opt., 1, 635.
\item Gramotnev, D. K., and Pile, D. F. P., 2000,
Opt. Quantum Electron., 32, 1097.
\item Nieminen, T. A., and Gramotnev, D. K., 2001, Opt. Commun., 189, 175.
\item Gramotnev, D. K., 2000, Frequency response of extremely
asymmetrical scattering of
electromagnetic waves in periodic gratings, 2000 Diffractive Optics
and Micro-Optics (DOMO-2000), Quebec City, Canada, pp.165--167.
\item Gramotnev, D. K., 2001, Opt. Quantum Electron., 33, 253.
\item Moharam, M. G., Grann, E. B., Pommet, D. A., and Gaylord, T. K.,
1995, J. Opt.
Soc. Am., A12, 1068.
\item Moharam, M. G., Pommet, D. A., Grann, E. B., and
Gaylord, T. K., 1995, J. Opt.
Soc. Am., A12, 1077.
\item Chateau, N., and Hugonin, J. P., 1994, J. Opt. Soc. Am., A11, 1321.
\item Li, L., 1996, J. Opt. Soc. Am., A13, 1024.
\item Liu, J., Chen, R. T., Davies, B. M., and Li, L., 1990,
Appl. Op., 38, 6981.
\item Li, L., Chandezon, J., Granet, G., and Plumey, J.  P., 1999,
Appl. Opt., 38, 304.
\item Akhmediev, N., and Ankiwicz, A., 1997, Solitons,
Non-linear Pulses and Beams
(London: Chapman \& Hall).
\item Gramotnev, D. K., and Pile, D. F. P., 2001,
J. Opt. A: Pure Appl. Opt., 3, 103.
\item Gaylord, T. K., and Moharam, M. G., 1985, Proc. IEEE, 73, 894.
\item Popov, E., and Mashev, L., 1985, Opt. Acta, 32, 265.
\item Biehlig, W., Hehl, K., Langbein, U., and Leberer, F.,
1986, Opt. Quantum Electron., 18, 219.
\item Biehlig W., 1986, Opt. Quantum Electron., 18, 229.
\item Shigesawa, H., and Tsuji, M., 1986, IEEE Trans.
Microwave Theory Tech., MMT-34, 205.
\item Borsboom, P.  P., and Frankena, H. J., 1995,
J. Opt. Soc. Am., A12, 1134.
\item Gramotnev, D. K., and Andres, Ch., Grazing-angle scattering of
waves in infinitely
wide periodic gratings, 2002, (unpublished).
\end{enumerate}

\end{document}